# Advantages of multi-dimensional biasing in accelerated dynamics: application to the calculation of the acid $pK_a$ for acetic acid

Jiasen Guo[1], Alberto Albesa[1,2], and Carlos Wexler[1]

[1]Department of Physics and Astronomy, University of Missouri, United States
[2]INIFTA, Universidad Nacional de La Plata, Argentina

**Abstract**

The use of accelerated sampling methods such as *metadynamics* has shown significant advantage in calculations that involve infrequent events, which would otherwise require sampling a prohibitive number of configurations to determine, e.g., difference in free energies between two or more chemically distinct states such as in the calculation of acid dissociation constants $K_a$. In this case, the most common method is to *bias* the system via a single collective variable (CV) representing the *coordination number* of the proton donor group, which yields results in reasonable agreement with experiments. Here we study the deprotonation of acetic acid using the reactive force field ReaxFF and observe a significant dependence of $K_a$ on the simulation box size when biasing only the coordination number CV, which is due to incomplete sampling of the deprotonated state for small simulation systems, and inefficient sampling for larger ones. Incorporating a second CV representing the distance between the $H_3O^+$ cation and the acetate anion results in a substantially more efficient sampling both accelerating the dynamics and virtually eliminating the computational box size dependence.

## 1. Introduction

The acid dissociation constant ($K_a$, or its potential $pK_a$ = -$\log_{10}(K_a)$) is a quantitative measure of the strength of an acid in solution; it is one of the most widely used material properties in the understanding of chemical and biochemical processes[1–4]. It is related to the difference in Gibbs free energy $\Delta G$ of the relation of interest by

$$pK_a = \frac{\Delta G}{\ln 10 \, RT}, \tag{1}$$

in this case the free energy difference between the protonated and deprotonated states of an acid. Many approaches have been used for the experimental determination of $pK_a$ values, including nuclear magnetic resonance, calorimetry, potentiometry, etc[5].

Generally, methods for the calculation of $\Delta G$ can be divided into three main categories depending on how the solvent is modeled: (i) explicit methods, (ii) implicit methods, and (iii) hybrid methods.

In explicit methods solvent molecules are modeled explicitly at various quantum mechanical or semiclassical levels. While the large number of solvent molecules often forbids the use of high-level *ab initio* methods, the adoption of Molecular Dynamics (MD) and Monte Carlo (MC) with various ways to account for quantum mechanical effects are thus employed. Although these methods are, in principle, more accurate than the others described below, the computational cost is extremely high especially for free energy calculations that require many states to be sampled.

In implicit solvent methods[6,7], (e.g., the Polarized Continuum Model (PCM), or the Conductor-like Screening Model (COSMO)) the solvent molecules are not included in the calculations/simulations directly, instead these are accounted typically by introducing a continuum medium with a dielectric constant that represents the screening of the electric fields due to the solvent. Additionally, the solutes reside in a cavity that is inaccessible by the continuous solvent, and the determination of this "bubble" is one of the most critical aspects of this method. While the concept of a dielectric constant itself is a reasonable approximation on long-range interactions, it is questionable at short distances, thus becoming ill-defined when local structures of the medium become nontrivial, e. g. the important hydrogen bonding effect in aqueous media (crucial for stabilizing hydrophilic molecules) [6,8–10]. In thermodynamic cycle methods isodesmic reactions are often used together with implicit solvent methods for the calculation of $\Delta G$[11] to partially compensate the inherent inaccuracies of the model.

Hybrid solvent methods reside in between the implicit and explicit methods and take into consideration the short-range solvent effect by including explicitly a small shell of solvent molecules near the solutes, while representing the remaining solvent by a continuous polarizable model[8]. In some cases, adding this explicit solvent molecule layer improve the prediction of $\Delta G$ (thus $pK_a$) considerably[12,13], however there is no clear way to determine the optimal number of explicit solvent molecules needed[13,14]. Furthermore, these added explicit solvent molecules are usually over-polarized and their entropic effect is often ignored[15]. The Quantum Mechanics and Molecular Mechanics method (QM/MM)[15,16] can also be regarded as a hybrid solvent method, where the solute molecules and the closet several solvation layers are treated at a high quantum chemistry level in the inner layer (primary system), and MM is used for the surrounding remaining explicit outer layers, (the secondary system). The interactions between the primary system and the secondary system then couples the two parts together[17]. However, MD simulations in aqueous environment require a rather complicated adaptive QM/MM scheme that allows the solvent molecules to diffuse through the boundary between the primary and secondary systems[14,17,18]. Additionally, for the calculation of the $pK_a$, one also needs to implement a method for locating extra protons in bulk water to allow the reconstruction of multiple primary system dynamically[19] which renders hybrid models hard to implement[20].

Reconstructions of the free energy landscape from MD simulations are commonly assisted by the use of enhanced sampling methods, such as metadynamics[21-24], umbrella sampling[25], or the adaptive biasing method[14,18,19]. These enhanced sampling methods help the simulated system escape from free energy minima and explore the configuration space that is otherwise inaccessible in normal MD simulations. To accomplish this, collective variables[26] (CV), which are reduced representations of the $3N$ Cartesian coordinates of the system ($N$ = number of atoms), are introduced to represent the state of the system in the characteristic chemical or physical process (e.g., the coordination number of an atom or ion). By biasing the CV, the simulation can often explore sufficiently the CV space even when large barriers (vs. the thermal energy $k_BT$) exist. This permits the calculation of free energy differences $\Delta G$ between various local minima representing the various configurations of interest, and the activation barrier along a transition path between them. Several publications have demonstrated the power of MD simulations assisted with enhanced sampling for the determination of the *pKₐ* values at various theoretical levels such as Car-Parrinello MD (CPMD)[10,27,28], Density Functional Tight Binding (DFTB)[29] and Reactive Force Field (ReaxFF)[30].

Results from previous studies with CPMD in conjunction with metadynamics[10,27,28] have shown good agreement with experimental values for *pKₐ*, however they were limited to only 52 to 60 water molecules, i.e., approximately three solvation shells. Meanwhile, only the coordination number of the proton donor group was used as the CV to guide the metadynamics sampling. Park *et al.*[31] pointed out that in addition to the coordination number of acetate oxygen atoms, two additional CVs could be used to characterize the acetic acid dissociation mechanism. These additional CVs describe the solvation structure of the hydronium, and the distance in between the hydronium and the acetate anion, respectively. Conceptually, these extra CVs allow a more complete characterization of the protonation/deprotonation pathways, and biasing them through metadynamics could permit the simulation to explore the deprotonated state more thoroughly, since any configurations with the coordination number of the acetate oxygen atoms below a certain cutoff value could be possibly regarded as the deprotonated state, regardless of the distance in between the hydronium and the acetate anion as well as how well the hydronium is solvated.

We therefore believe it is of significant practical interest to determine if the calculated *pKₐ* depends on how the configuration space is sampled. Whereas other studies have focused on the accurate determination of *pKₐ* for various substances, we aim to gain insight on how these higher-dimensional free energy landscapes (depending on the various CVs) affect the calculation of *pKₐ*. We thus employ the reactive force field ReaxFF[32] for our calculations. The ReaxFF is substantially faster than fully QM methods while still allowing bond formation and braking, unlike the traditional MM force fields, thus enabling the simulation of chemical reactions.

This article is organized as follows. In **Section 2**, we present the computational methodology specific to this study. The results of the simulations are presented in **Section 3**, and the consequences are summarized and discussed in **Section 4**. We found that by both biasing the distance between the acetate anion and the hydronium together with the acetate oxygen coordination number, sampling of the full configuration space can be accelerated resulting in accurate, system size-independent predictions of the acetic acid *pKₐ*. This not only allows faster exploration of large simulation systems but improved accuracy of more modest ones. Our results thus open a new venue for the accurate prediction of various properties like the acid *pKₐ* at higher confidence levels with modest computational cost.

## 2. Methodology

The LAMMPS simulation package was used to perform MD simulations[33]. The ReaxFF potential developed by Sengul et al.[30] was used to describe the interactions in the simulated system. This potential was developed based on the water potentials from van Duin et al.[34] as well as a potential parameterized for the modeling of amino acid and small proteins[35] derived from an earlier glycine potential[36]. It has been successfully applied for the modeling of the solvated structure and transport properties of water, hydronium and hydroxide[37,38,39].

Initially the simulated system consists of one acetic acid molecule solvated by a certain number of water molecules in a cubic box. Two different simulation boxes containing $N_w$ = 60 and 200 water molecules respectively were used in order to study the dependence of the calculated $\Delta G$ of the acetic acid deprotonation process on the simulation box size. For each simulation box, a 300 ps long NPT MD simulation with a Nose-Hoover thermostat and barostat was performed first to relax the simulation box, followed by a NVT MD simulation with a Nose-Hoover thermostat in place to equilibrate the simulated system. Although reproducing the correct water density is obviously critical for the determination of $pK_a$,[10] enforcing this density may result in systems with extreme pressures and artificially modified water molecule mobilities inconsistent with the force field used, which might affect the proton transfer in bulk water via the Grotthuss mechanism[40]. Therefore, we relaxed the simulation box dimensions constrained by symmetry in the NPT portion of the simulation so that the simulations are self-consistent. The resulting density of the water ranges from $\rho_w$ = 0.85 to 0.91 g/cm$^3$ (pure water, no acetic acid included). Configurations after equilibration of 425 ps, 450 ps, 475 ps, 500 ps, 525 ps were then taken as the initial configurations of the metadynamics simulations that followed.

All MD simulations were performed at 300 K with a time step of 0.25 fs. A well-tempered metadynamics algorithm[23] with adaptive Gaussians[24] implemented in Plumed[41] was used to enhance the sampling of the CV space. Unless explicitly noted, the initial Gaussian height was set at 0.1 kcal/mol with a bias factor of 15. Multivariate Gaussian biasing potentials were deposited in the CV space every 400 steps (100 fs) with the width determined using the last 100 steps (25 fs). The well-tempered scheme was adapted to avoid irreversible transition to uninterested region in the collective variable space due to over-depositing biasing Gaussian potential[23]. Note that this advantage of the well-tempered scheme is particularly important here due to the dynamical feature of the CV adapted.

Two CVs were used in this work. The first reflects the coordination number of the acetate oxygen atoms (*Coordination* hereafter), given by

$$n_X = \sum_{i=1}^{N_H} \frac{1 - \left(\frac{r_{OH_i}}{r_c}\right)^{12}}{1 - \left(\frac{r_{XH_i}}{r_c}\right)^{24}}, \qquad (2)$$

where $X$ stands for either one of the two acetate oxygen atoms, and $N_H$ is the number of protons in the simulation box excluding those belonging to the methyl group. Furthermore, $r_{OH}$ is the distance between the acetate oxygen atoms and a proton, and $r_c$ denotes a cutoff distance. This *Coordination* CV has been widely used in the literature to describe the acid dissociation process[10,27,28,31]. In this work, we found it convenient to have a sharper discrimination between the protonated and the deprotonated states and thus used 12 and 24 powers instead of the more common 6 and 12. The cutoff distance was set at $r_c = 1.27$ Å.

The second CV was adapted from Ref. 31, which effectively describes the distance between the acetate anion and the hydronium (*Hydrodist* hereafter):

$$R = f(n_{OA}) \left[ \frac{\sum_{\substack{i \in O_W \\ k \in O_A}} r_{ik} \exp[\lambda h(n_i)]}{2 \sum_{i \in O_W} \exp[\lambda h(n_i)]} \right], \quad (3)$$

where $R$ is the distance from the hydronium oxygen atom to the center of mass of the two acetate oxygen atoms, $O_W$ and $O_A$ denote the water oxygen and acetate oxygen atoms, respectively, $n_i$ and $n_{OA}$ are given by **Eq. (2)** and $\lambda$ is a parameter which controls the sensitivity of the *Hydrodist* CV on the coordination number of individual water oxygen atoms (see **Appendix A** for more details). Note that a cutoff function $f(n_{OA})$ correlates the two CVs; this cutoff function is also defined as a rational function, such that the *Hydrodist* CV is meaningful only when the *Coordination* CV drops below a certain cutoff. It is worth noticing that a relatively small change in the CVs defined above may correspond, indeed, to a significant solvent rearrangement (i.e., a significant change in the system's microstate), suggesting that a low metadynamics depositing rate is required[23]. At the same time, this means that a long simulation time is necessary for the system to revisit the protonated state and the deprotonated state multiple times in order to obtain statistically meaningful averages. Therefore, the well-tempered scheme of the metadynamics algorithm was used with a fine tunned bias factor, such that long time simulations without over-depositing biasing potential can be achieved.

Free energy landscapes were constructed using the estimator of Eq. 18 in Ref. 24 using Plumed[41]; the convergence of each free energy landscape was determined by monitoring the fluctuation of the free energy difference between the protonated state and the deprotonated state. VMD was used for the visualizations of atomic configurations[42].

## 3. Results and Discussion

Metadynamics with a single CV (*Coordination*):

To see the effect of the simulated system size on $\Delta G$ between the protonated state and the deprotonated state, independent simulations with $N_W = 60$ and 200, biasing only the *Coordination* CV (**Eq. 2**) were initially performed. For $N_W = 60, 200$ the cubic simulation box size was $L \approx 12.7$, 19.0 Å, respectively. **Figure 1** compares the cumulative free energy surface (FES) per 1,000 Gaussian potentials as well as the corresponding CV trajectories from two metadynamics runs. In order to demonstrate the temporal variation of the cumulative FES clearly, each simulation was performed long enough such that at least three deprotonation events were observed.

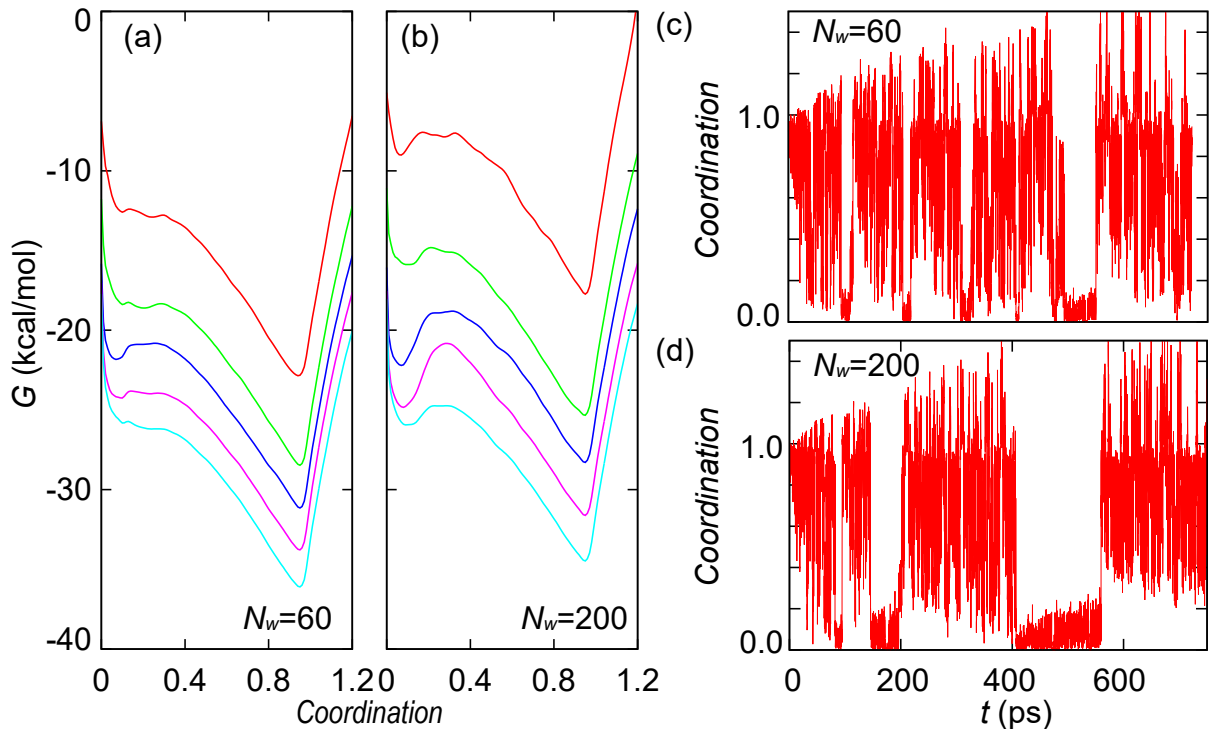

**Figure 1. Cumulative free energy surface (FES) and corresponding *Coordination* CV trajectories during metadynamics simulations of simulated systems of $N_W$ = 60, 200 and cubic simulation cells with $L \approx 12.7$, 19.0 Å, respectively. (a)** and **(b)** Cumulative FES of the simulated system of $N_W$ = 60 and 200, respectively, plotted every 1,000 Gaussian potentials. **(c)** and **(d)** Trajectories of the *Coordination* CV as a function of simulation time $t$ during the respective metadynamics simulations.

A well-defined free energy minimum associated with the protonated state is shown by both simulations (i.e., deep minimum and independent of each individual run). However, a well-defined minimum in the deprotonated state was not always found. For the small system with $N_W$ = 60, $L \approx$ 12.7 Å, the deprotonated state is very flat, indicating a relatively unstable state which is unphysical. In the larger system $N_W$ = 200, $L \approx$ 19.0 Å there is (in most cases) a well-developed minimum, thus indicating a relatively stable deprotonated state and making it longer-lived. However, this also slows down the rate of protonation-deprotonation cycles significantly which results in higher variability and lower quality statistics. In fact, one of the most undesirable consequences is the variability of the protonation-deprotonation times, especially for the larger system, as seen in **Fig. 1d**.

This can be also observed in **Fig. 2** which, without loss of generality, presents the time series of $\Delta G$ per 1,000 Gaussian potentials of several independent runs. Initial configurations were taken from previous NVT simulations equilibrated by varying simulation lengths as labelled. It is observed that with the increase of the simulated system size (larger $N_W$, $L$), the $\Delta G$ of individual runs (data sets with the same label) and for distinct runs vary more dramatically. As stated above, the large variability of protonation-deprotonation times, and variations of calculated $\Delta G$ imply that one would have to run numerous independent simulations and calculate the average $\Delta G$ as the best estimate.

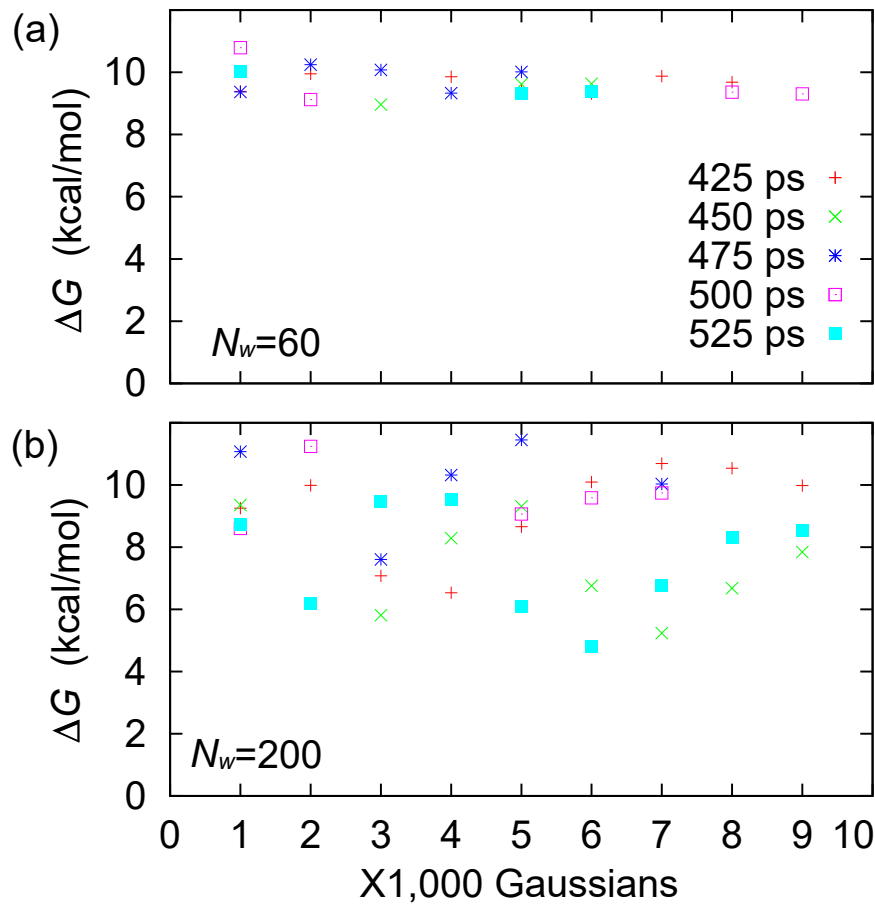

**Figure 2. Time series of $\Delta G$ between the protonated state and the deprotonated state of independent runs of two different system sizes. (a)** Simulations with $N_W = 60$. **(b)** Simulations with $N_W = 200$. The legends denote the equilibration lengths before applying metadynamics. $\Delta G$ were calculated every 1,000 Gaussian potentials. Missed data points indicate the lack of a well-defined minimum in the deprotonated state.

**Fig. 1** and **2** together demonstrate that biasing only the *Coordination* CV results in a significant dependence of $\Delta G$ on the simulated system size. The predicted $\Delta G \approx 10.7$ kJ/mol, corresponding to a $pK_a \approx 7.3$ for the small system is quite distinct from the experimental value 4.76. An average $pK_a \approx 5.1$ was obtained for the larger system, but with substantial variations between runs, as described above.

It is easy to understand that the failure of the prediction of acetic acid $pK_a$ in the small system is due to inadequate sampling of the configurations where the hydronium and the acetate anion are independently well-solvated and separated from each other. We observed that for the hydronium and acetate anion to be solvated independently, the minimum separation distance is roughly $d_{\text{full solvation}} \approx 5.6$ Å (see **Appendix B**). Additionally, note that the "escape distance" from the Coulomb interaction in a medium (i.e., the Bjerrum length, $\lambda_B$) is such that the interaction potential is smaller than the thermal energy: $e^2/(4\pi\epsilon_r\epsilon_0 \lambda_B) < k_B T$, which for water at room temperature is $\lambda_B \approx 7$ Å[43]. In a cubic simulation box with periodic boundary conditions the maximum separation between acetate anion and the hydronium is $d_{\max} = \sqrt{3}/2\, L$ at the "cube corners" which is large enough ($d_{\max} = 10.9$ Å), but there is a relatively small volume available in those areas, making it hard to sample regions where the counterions are fully solvated effectively, reflected by the shallow or non-existent free energy minima for the deprotonated states, and an inaccurate $\Delta G$. It is easy to see that a larger system has a higher probability to sample fully separated counterions, but the long protonation-deprotonation cycles and their stochastic nature makes for correct-in-principle but rather variable $\Delta G$. For more details see **Appendix C**.

Below we discuss how the multi-dimensional metadynamics procedure proposed here can sample efficiently even the small "cube corners" of the small system while also improving the sampling efficiency of large systems. Thus, we obtain a more predictable, system-size independent calculation of $\Delta G$ that works well for both relatively small and large system sizes.

Multi-dimensional metadynamics (adding the *Hydrodist* CV):

Following the work of Park *et al.*[31], *Hydrodist* (**Eq. 3**) was used as a second CV in this work. **Figure 3a** shows a 2D FES of the acetic acid deprotonation process and visualizations of the protonated (**Fig. 3b**), contact-ion (**Fig. 3c**), and deprotonated (**Fig. 3d**) states. The metadynamics simulation was carried out in a system of 200 water molecules and started from an initial configuration equilibrated for 425 ps. Initially the system started from the protonated state, where the charge-neutral acetic acid was well-solvated by the surrounding water molecules. As more and more Gaussian potentials were deposited along the *Coordination* CV, the contact ion state occurred as a transition in between the protonated state and the deprotonated state. In this contact ion state, a proton bridges together the acetate anion and the approaching water molecule; meanwhile, the proton is closer to the approaching water molecule than to the acetate anion, leaving the approaching water molecule a hydronium hydrogen-bonding to the acetate oxygen atoms. One can see that the first solvation layer of the so-formed hydronium is completed with the help of the acetate oxygen atoms. The distance between the acetate oxygen atom and the bridging proton is around 1.4 Å, in good agreement with the value in Ref. 10. The hydronium then travels away from the acetate anion and the first deprotonation event occurs after 70 ps of depositing Gaussian potentials.

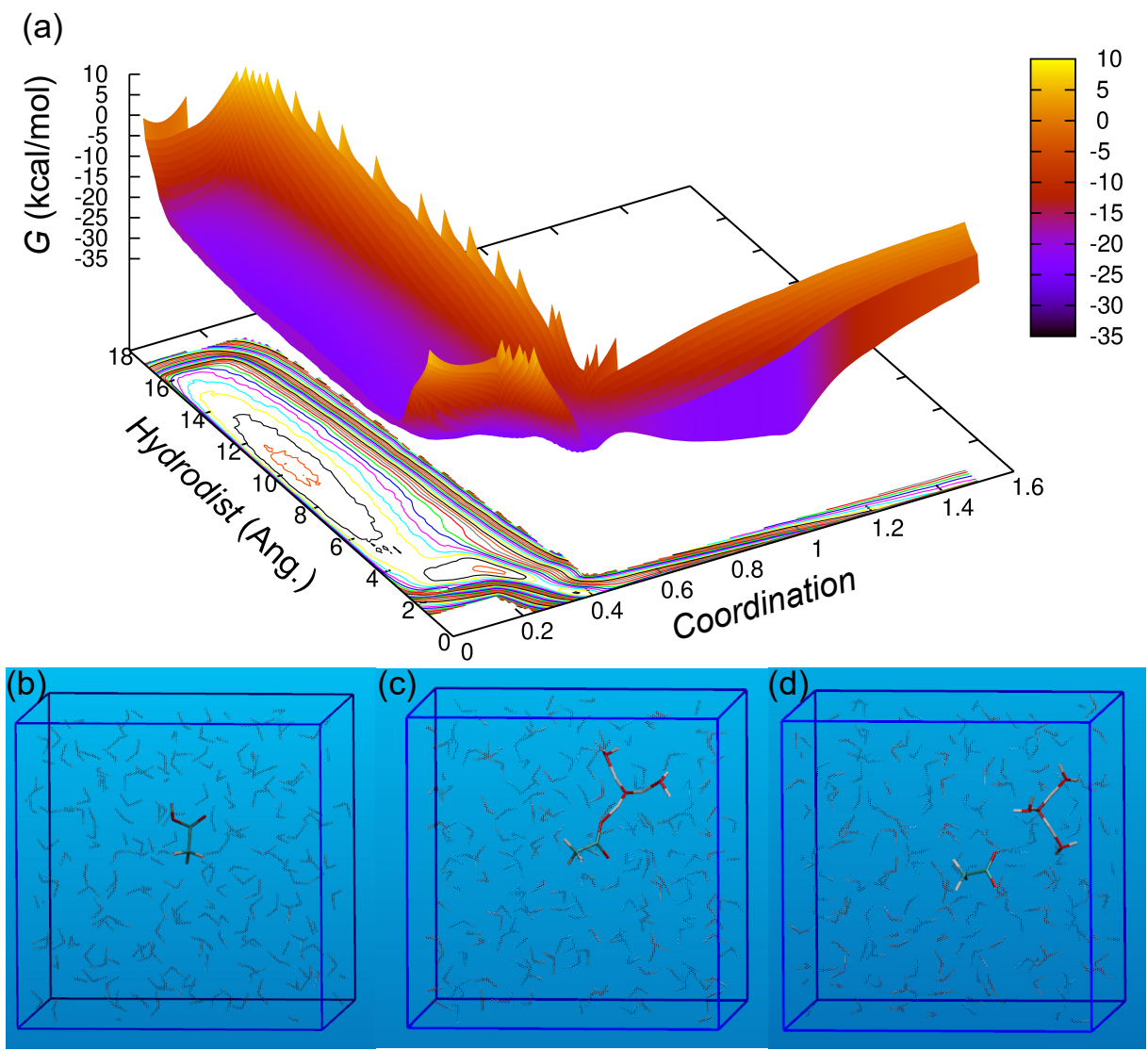

**Figure 3. Visualization of the the deprotonation process of acetic acid as well as the corresponding FES.** **(a)** 2D FES of acetic acid deprotonation process. the protonated state is reproduced as a deep minimum at around (0.95, 0.0), while the deprotonated state is represented by a broad flat region from around (0.1, 5.0) to around (0.1, 15.0). Another local minimum appearing at around (0.2, 3.0) corresponds to the contact-ion state on the path in between the protonated state and the deprotonated state. $\Delta G$ between the protonated state and the broad flat region is about 8 kcal/mol; the $\Delta G$ between the protonated state and the contact-ion state is 8 kcal/mol. **(b)** The protonated state. The charge-neutral acetic acid molecule is solvated in bulk water. **(c)** The contact-ion state. An extra proton is close to a water molecule near the acetate anion, making the water molecule essentially a hydronium whose first solvation layer is completed by two neighboring water molecules and an acetate oxygen atom. **(d)** The deprotonated state. The hydronium and the acetate anion are solvated independently and relatively far from each other.

A typical configuration of the deprotonated state is shown in **Fig. 3d**. Both the hydronium and acetate anion are solvated independently and away from each other. Schematic diagrams of the deprotonation process are also given in **Appendix B**. It is easy to see (**Fig. 3a**) that the protonated state is represented by a deep minimum located at around (*Coordination*, *Hydrodist*) ≈ (0.95, 0.0). This free energy minimum is very narrowly confined around *Hydrodist* = 0, which is expected since in absence of deprotonation (large *Coordination*) the switching function $f(n_{OA})$ (**Eq. 3**, see also **Appendix A**) guarantees *Hydrodist* ≈ 0. On the other hand, the deprotonated state is found to be a wide flat region along the (*Coordination*, *Hydrodist*) ranging approximately (0.1, 5.0) to (0.1, 15.0). Since the *Hydrodist* collective variable describes the distance in between the hydronium and acetate anion, and the change of this distance is invoked by proton transfer among bulk water molecules, the FES along the *Hydrodist* collective variable shall reflect the energetics of the proton transfer in bulk water. It has been observed in classical MD that the free energy barrier for an extra proton to transfer in between two neighboring water molecules is ~ $k_BT$, and is even reduced to only a fraction of $k_BT$ by the quantum mechanical zero-point energy of the nuclei in *ab initio* MD[44,45] and QM/MM MD[46]. Therefore, such a flat FES in the deprotonated state is not surprising. The minimum at around (0.2, 3.0) corresponds to the contact ion state.

Recall that 1D metadynamics simulations (**Fig. 2**) reveal large temporal and inter-run variations of ∆$G$ as system size increases. Here we show that this dependence between ∆$G$ variability and the system size can be explained by the inherited ineffectiveness of the *Coordination* CV. In cases when only the *Coordination* number is biased, once the deprotonation proceeds beyond the contact-ion state, one loses the control on how long the extra proton will stay in bulk water before reprotonation. This is because for the reprotonation to happen, it is energetically not likely for the acetate anion to grab a proton from a neighboring charge-neutral water molecule hydrogen-bonding to itself, which will otherwise leave a hydroxyl group. Instead, the extra proton in the bulk water must migrate to this water molecule via proton transfer following the Grotthus mechanism[47] and form the contact-ion state (**Fig. 3c,**), such that the extra proton in the contact-ion state can be captured by the acetate anion.

Since the rational function used in the definition of the *Coordination* CV decays rapidly, the propagation of the biasing force due to the presence of the already-deposited Gaussian potentials to the most majority of the protons in the simulated system is negligible, only the closet protons will feel the biasing force. This means that once the distance between the extra proton and acetate anion goes beyond the effective range of the rational function, there will be practically no biasing force driving the proton transfer making the biasing itself very ineffective for the deprotonated state. A milder rational function with a large cutoff distance[10] (**Eq. 2**) in principle would increase the effective range of the biasing potentials. However, our previous simulation experiences using the same ReaxFF and CPMD suggest that a mild rational function with a large cutoff distance tends to overestimate the coordination number due to the dynamically-evolving hydrogen bonds between the acetic acid (or acetate anion) and the surrounding water molecules, making the protonated state and the deprotonated state indistinguishable from the perspective of the *Coordination* CV. Consider the stochastic nature of proton transfer in bulk water: one can see that in simulations biasing only the *Coordination* CV, reprotonation occurs only when the extra proton diffuses towards the acetate anion to form the contact-ion. Thus, the time that the system will reside in the deprotonated state before reprotonation directly relates to the free space inside the simulation box available for the hydronium and the acetate anion to be independently-solvated. This free space grows as the simulated system size increases, hence on average a deeper free energy minimum associated with the deprotonated state. However, it is possible that in some deprotonation-reprotonation cycles, the system resides in the deprotonated state for a long time, while in other cycles, it quickly returns to the protonated state (**Fig. 1**). Therefore, in bigger systems a substantial temporal variation of the $\Delta G$ of individual runs are expected and the same argument also explains the wider distribution of the $\Delta G$ among independent runs performed in such a system (**Fig. 2**).

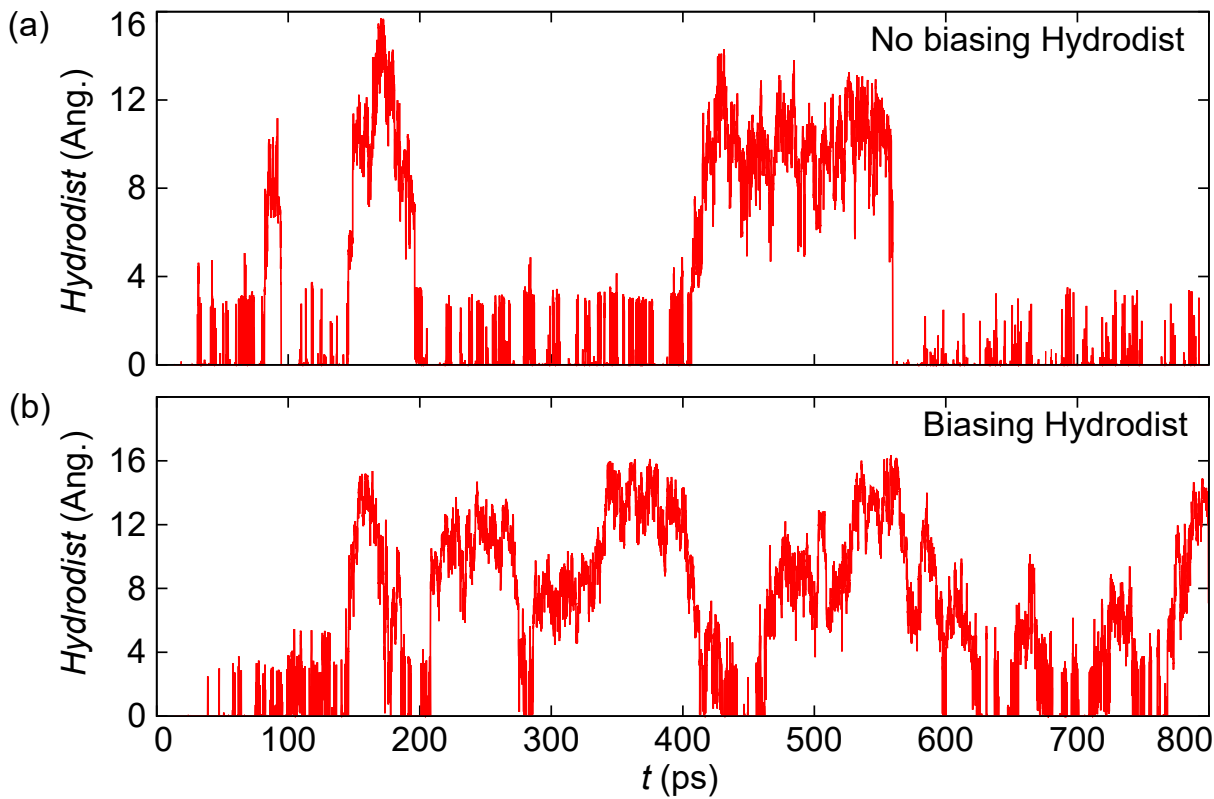

**Figure 4. Trajectories of the *Hydrodist* CV with and without *Hydrodist* medadynamics biasing. (a)** Trajectory post-processed from a 1D metadynamics simulation where only the *Coordination* was biased. **(b)** Trajectory when both *Coordination* and *Hydrodist* were biased (with initial Gaussian height 0.05 kcal/mol and bias factor 20). One can see that biasing the *Hydrodist* CV helps to explore the CV space more comprehensively. In both cases, metadynamics simulations were performed starting from an initial configuration equilibrated for 525 ps.

The incorporation of the *Hydrodist* CV in metadynamics simulations helps to bias the proton transfer. Effectively, the *Hydrodist* CV helps drive the extra proton in between the vicinity of the acetate anion and the edges of the simulation box and force the system to explore all the possible configurations in the deprotonated state. This also ensures that multiple deprotonation-reprotonation cycles can be achieved in a controllable manner. **Figure 4** shows comparisons of the trajectories of the *Hydrodist* CV from an 1D (biasing only the *Coordination*, *Hydrodist* was calculated via post processing) and a 2D (basing both *Coordination* and *Hydrodist*) metadynamics simulations. One can see that when both CV are biased, the CV space along *Hydrodist* is sufficiently explored.

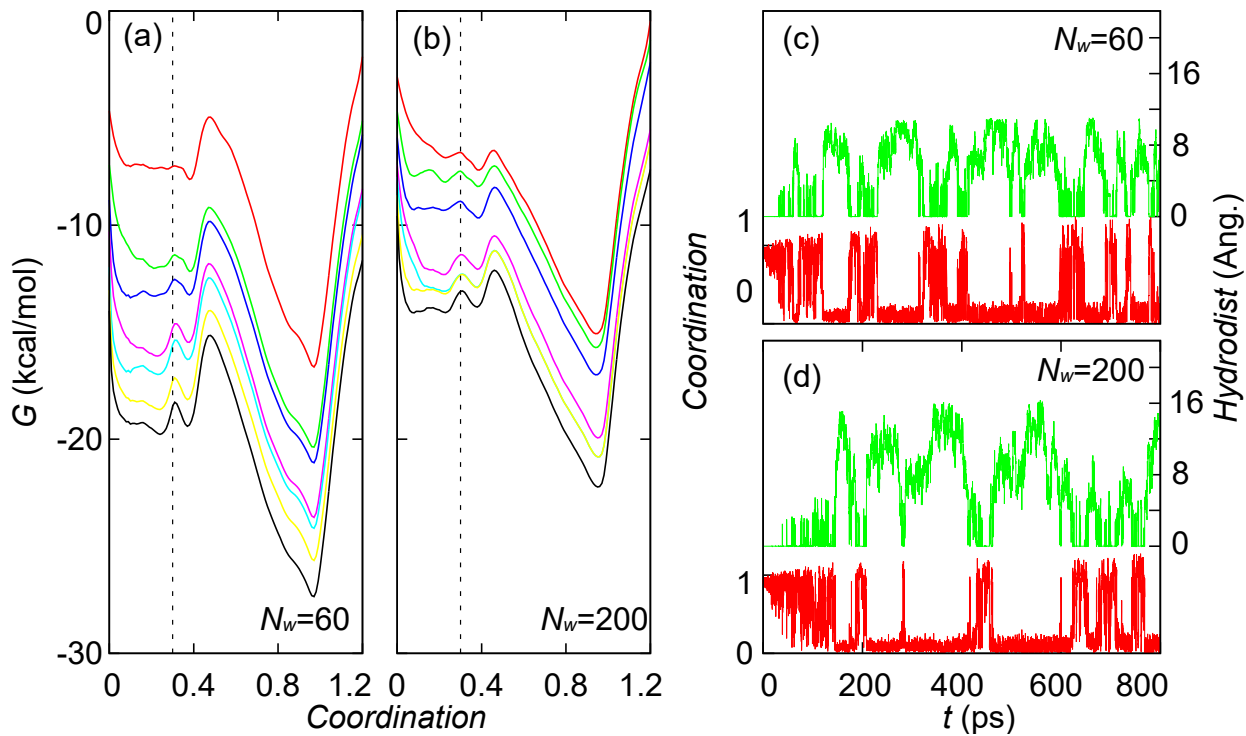

**Figure 5. Cumulative FES and trajectories of the CV when biasing both *Coordination* and *Hydrodist* CVs** during metadynamics simulations of simulated systems of different sizes. **(a)** and **(b)** Cumulative FES of the simulated system of $N_W = 60$ and 200, respectively, plotted every 1,000 Gaussian potentials. Metadynamics simulations were performed with system sizes as labeled, starting from initial configurations equilibrated for 525 ps. The deprotonated state on the presented 1D FES, which can be attributed to the broad flat region and the contact ion state on the 2D FES, is well-reproduced in the range from 0.0 to 0.3 (Dash line to the left) regardless of the simulated system size. **(c)** and **(d)** Trajectories of the CV as a function of simulation time during the respective metadynamics simulations. Compared to **Fig. 1**, the trajectories show faster and more predictable deprotonation-reprotonation cycles.

In order to demonstrate, comparatively, the effect of the *Hydrodist* CV on the reconstructed FES, **Fig. 5** presents the cumulative FES and CV trajectories of two metadynamics simulations performed with 60 and 200 molecules, respectively, biasing both the *Coordiantion* and *Hydrodist* CVs. The presented 1D FES were obtained by integrating out the *Hydrodist* CV. Clearly, the free energy minimum corresponding to the deprotonated state is always well-defined from *Coordiantion* $\in [0.0, 0.3]$ regardless of the simulated system size. This free energy minimum is contributed from both the broad flat region and the contact-ion state on the 2D FES (see **Fig. 3a**) and has manifested itself by the two growing minima within this range. The reason for the minimum at around 0.4 is not clear to us, however it might be an artifact of the switching function included in the definition of the *Hydrodist* CV (see **Appendix A**). Overall, the shapes of the cumulative FES resemble each other quite well regardless of the simulated system size. A comparison between the FES of **Fig. 1** and **5** shows that incorporating *Hydrodist* to the metadynamics biasing results in faster, more regular protonation-deprotonation cycles, achieved by the forced exploration of the deprotonated state, which enables a more efficient sampling of the relatively small configuration spaces of the "cube corners" of the simulation boxes, resulting in less dependence of Δ$G$ on system size, as explained earlier.

**Figure 6** shows the time series of the free energy difference between the protonated state and the broad flat region (Δ$G$* hereafter), as well as Δ$G$ of independent runs performed with simulated system of 60 and 200 water molecules, respectively. After depositing roughly 20,000 Gaussian potentials, Δ$G$* converges to a range from 8.0 to 10.0 kcal/mol in both simulated systems. Integrating out the *Hydrodist* CV further lowers the free energy difference by about 2 kcal/mol. Thus Δ$G$ converges to 6.0 to 7.5 kcal/mol. A comparison to **Fig. 2** shows that the incorporation of the *Hydrodist* CV in metadynamics simulations, reduces or even eliminates the system size dependence of Δ$G$.

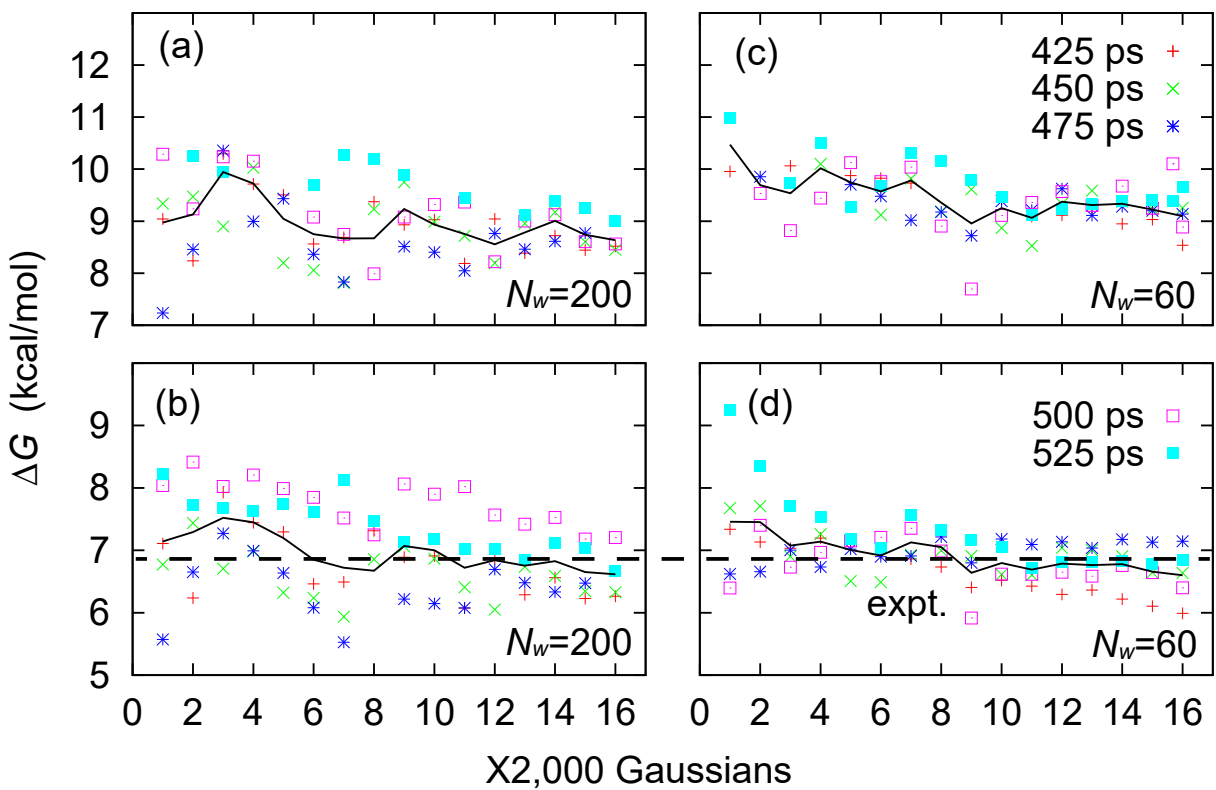

**Figure 6. Time evolution of the free energy difference between the protonated state and the broad flat region (Δ$G$*) and Δ$G$ simulated with system of different sizes. (a)** and **(b)** Δ$G$* and Δ$G$ of independent runs simulated with $N_W$ = 200. **(c)** and **(d)** Δ$G$* and Δ$G$ of independent runs simulated with $N_W$ = 60. The dimension reduction is performed by integrating out the *Hydrodist* CV. In both cases, Δ$G$* converges to 8.0 to 10.0 kcal/mol, and Δ$G$ converges to a range from 6.0 to 7.5 kcal/mol after 20,000 Gaussian potentials. Solid lines denote the average value over independent runs, dash line denotes the experimental Δ$G$.

With a reasonably size- and initial configuration-independent $\Delta G$, we compute the average from 69 data points after 20,000 Gaussian potentials: $\Delta G = 6.8 \pm 0.4$ kcal/mol. Application of **Eq. 1** yields $pK_a = 4.9 \pm 0.3$ for acetic acid in reasonably good agreement with the experimental value $pK_a = 4.76$[48].

## 4. Conclusions

In summary, we have demonstrated the system-size dependence of the estimated free energy difference $\Delta G$ (or $pK_a$) when only the Coordination number is biased in metadynamics simulations. It is due to the insufficient sampling of the deprotonated state rooted in the stochastic nature of the proton transfer process in bulk water as well as the ineffective propagation of the biasing force. An additional degree of freedom (*Hydrodist*) representing the distance in between the hydronium and the acetate anion in the deprotonated state has been proven essential for a thorough sampling of the deprotonated state. Incorporation of *Hydrodist* renders the proton transfer in bulk water from stochastic to a bias force-driven process, leading to fast deprotonation-reprotonation cycles, and hence statistically more reliable estimates of $\Delta G$. The calculated best estimate for the acetic acid $pK_a$ of 4.9 (0.3) agrees well with the experimental value 4.76 regardless of the simulated system size. Thanks to its ability to virtually eliminate the system-size dependence of $\Delta G$, the multi-dimensional biasing strategy used in this work could be particularly useful when high-level *ab initio* methods are in demand with limited computational resources. In the meantime, it also provides valuable insights for the computational modeling of more complicated reactions where multiple proton donors and acceptors coexist.

## 5. Acklowledgements

This work was funded in part by the National Science Foundation Grant No. IIP-2044726.

# APPENDICES

## A. The *Hydrodist* collective variable

The *Hydrodist* CV adapted in this work is defined as

$$R = f(n_{OA}) \left[ \frac{\sum_{\substack{i \in O_W \\ k \in O_A}} r_{ik} \exp[\lambda h(n_i)]}{2 \sum_{i \in O_W} \exp[\lambda h(n_i)]} \right], \quad (A1)$$

here $f(n_{OA})$ denotes a switching function that correlate the *Hydrodist* and *Coordination* CV; $n_i$ is the coordination number of the $i^{th}$ water oxygen atom and $h(n_i)$ is a rational function that scale $n_i$ to the range $0 \sim 1$:

$$f(n_{OA}) = \frac{1 - \left(\frac{n_{OA}}{f_c}\right)^{12}}{1 - \left(\frac{n_{OA}}{f_c}\right)^{24}} \quad , \tag{A2}$$

where $n_{OA}$ is the *Coordination* CV, $f_c$ a selection cut off ($f_c = 0.3$ in this work), such that when a molecule is in the protonated state ($n_{OA} \approx 1$), the switching function $f(n_{OA}) \approx 0$, thus *Hydrodist* CV $\approx 0$, as well; when in the deprotonated state ($n_{OA} < 0.3$), the switching function $f(n_{OA}) \approx 1$, then the *Hydrodist* CV becomes defined in terms of the counterions distances. $n_i$ and $h(n_i)$ are given by:

$$n_i = \sum_{j=1}^{N_H} \frac{1 - \left(\frac{r_{O_i H_j}}{O_c}\right)^{12}}{1 - \left(\frac{r_{O_i H_j}}{O_c}\right)^{24}} \quad , \tag{A3}$$

$$h(n_i) = \frac{1 - \left(\frac{n_i}{s_c}\right)^{36}}{1 - \left(\frac{n_i}{s_c}\right)^{72}} \quad , \tag{A4}$$

where $r_{O_i H_j}$ is the distance between the $i^{\text{th}}$ water oxygen atom and the $j^{\text{th}}$ proton, $O_c$ the cutoff distance (here $O_c = 1.4$ Å) and $s_c$ a parameter to distinguish the hydronium from water molecules (here $s_c = 2.2$). With an appropriate control parameter $\lambda$, the weighted average in the definition of the *Hydrodist* CV effectively picks up the distance in between the center of mass of the two acetate oxygen atoms and the hydronium oxygen atom.

**B. Schematic diagram of the acetic acid deprotonation process**

**Figure A1** presents a schematic diagram of the deprotonated process assumed in this work. In the protonated state, the charge-neutral acetic acid is solvated by the surrounding water molecules (Scheme (A)). By biasing the coordination number of the acetate oxygens, a contact-ion state forms before the proton migrates into bulk water via proton transfer (Scheme (B)). For convenience in calculation, the dynamical O-H monovalent bond length and the hydrogen bond length can be roughly approximated to be 1.0 Å and 1.8 Å, respectively. In the contact-ion state, the *Hydrodist* CV, defined as the distance in between the center of mass of the two acetate oxygen atoms and the hydronium oxygen atom is roughly 2.8 Å. The deprotonation process then proceeds beyond the contact-ion state, and the hydronium starts to leave away from the acetate anion (Scheme (C)). However, at this stage the first solvation layer of the acetate anion and the hydronium still share one water molecule; the corresponding *Hydrodist* is roughly 4.9 Å. Farther away from each other, the acetate anion and the hydronium then have independent first solvation layers (Scheme (D)). The *Hydrodist* of such an independently solvated structure is roughly 5.6 Å, or above.

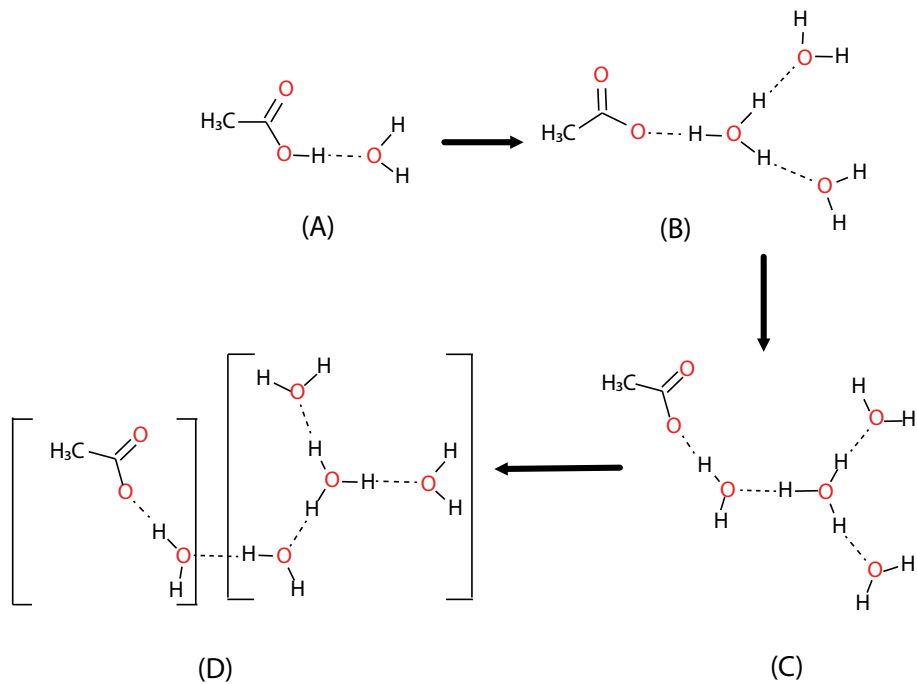

**Figure A1. Schematic diagram of the acetic acid deprotonation process.** (A) Protonated state, (B) contact ion state, (C) transition structure and (D) independently solvated structure. Valent bonds are represented by solid lines, hydrogen bonds are represented by dash lines.

## C. Graphical representation of the free space in boxes of different sizes

**Figure A2a** graphically represents the free space available for the hydronium and the acetate anion to be independently solvated (**Fig. A1 (D)**). The inner sphere, of radius 5.6 Å represents the volume centered around the acetate anion in which a hydronium would not be independently solvated (i.e., the volume for which the only possible configurations are those of **Fig. A1 (A-C)**. The inner cube of side $L = 12.6$ Å is the simulation box for 60 water molecules; the outer cube is the equivalent $L = 19.0$ Å for 200 water molecules. For each simulation, the volume between the cube and the sphere is the region for which the independently solvated structure exists (**Fig. A1 (D)**). As can be readily observed, even the small simulation box has regions for which the independently solvated configuration can be found, thus a simulation with only 60 water molecules should be able to express the independently solvated configuration. However, for the counterions to be effectively independent, i.e., with negligible Coulomb interaction between them, they should be separated at least by the Bjerrum length, $\lambda_B \approx 7$ Å[43]. **Figure A2b** shows that a scheme with the inner sphere radius equal to $\lambda_B$. It can be plainly seen that for simulations with only 60 water molecules, only relatively small patches near the cube diagonals satisfy this condition, thus statistically simulations that sample the volume of the cube may have difficulty exploring effectively the independently solvated configuration when the counterions are well-separated. We attribute this lack of proper sampling to the existence of the very shallow local minimum for *Coordination* CV ≈ 0 seen in **Fig. 1** and the incorrect resulting Δ*G* (**Fig. 1**, **2**) for metadynamics simulations biased exclusively with the *Coordination* CV. It is also evident that for the larger simulation box (200 water molecules) the volume available for this configuration is vast, thus readily explored, leading to deeper local minima, and a more accurate Δ*G*; however this vast space also makes random sampling of all configurations very slow due to the stochastic nature of proton transfer in bulk water, leading to the slow and unpredictable dynamics observed (**Fig. 1**, **2**).

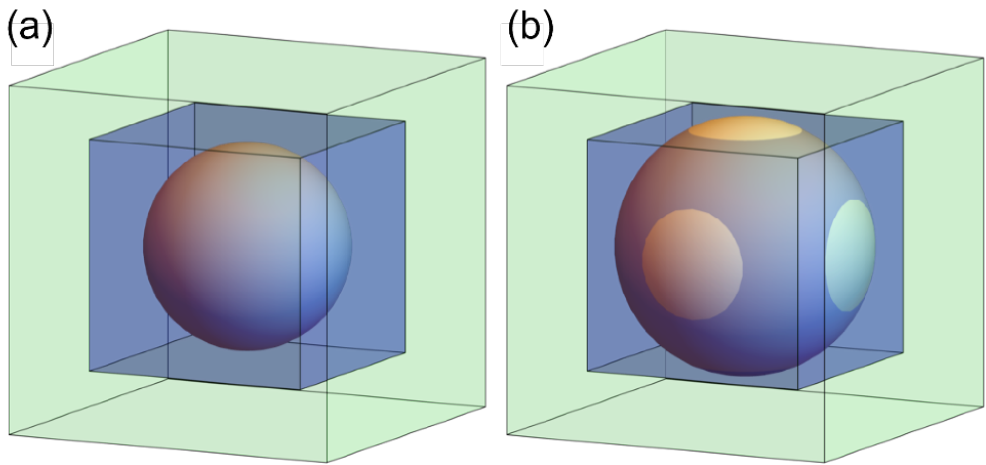

**Figure A2. A graphical representation of the space available for the independently solvated configuration.** The inner and outer cubes represent the simulation boxes with $L = 12.6$ and $19.0$ Å corresponding to simulations with 60 and 200 water molecules, respectively. The spheres represent **(a)** the minimum distance between the acetate and hydronium counterions for the independently solvated configuration ($R = 5.6$ Å, see **Fig. A1 (D)**), and **(b)** the Bjerrum length, $\lambda_B \approx 7$ Å, the minimum distance for which the Coulomb interaction becomes smaller than $k_B T$[43].

**Figure A3** compares the probability distributions of the *Hydrodist* CV (representing the distance between the acetate and hydronium counterions) for $N_W = 60$ and $200$ water molecules for metadynamics performed with biasing 1-dimensionally, i.e., using exclusively the *Coordination* CV, or 2-dimensionally, i.e., with biasing both the *Coordination* and *Hydrodist* CVs. The main observations are:

1) The large peak near the origin corresponds to the protonated state.
2) For both $N_W = 60$ and $200$, biasing the *Hydrodist* CV permits a more efficient sampling of large *Hydrodist* values.
3) In the absence of *Hydrodist* biasing, there is essentially no sampling of *Hydrodist* > 9 and 14 Å for $N_W = 60$ and $200$ ($L = 12.6$ and $19.0$ Å), corresponding to $R \approx 2^{-1/2} L$, i.e., for distances larger than the edges of the cube (see **Fig. A2**).
4) Once biasing on *Hydrodist* is added, the system explores effectively up to *Hydrodist* ≈ 11 and 16.5 Å for $N_W = 60$ and $200$ ($L = 12.6$ and $19.0$ Å), corresponding to $R \approx d_{max} = \sqrt{3}/2\ L$, i.e., the simulation can effectively explore all the way to the cube corners.
5) In absence of *Hydrodist* biasing there is a substantial peak around *Hydrodist* ≈ 3 Å, which we

attribute to the contact-ion state (**Fig. A1 (B)**), this configuration's apparent stability is not present with the 2-dimensional biasing.

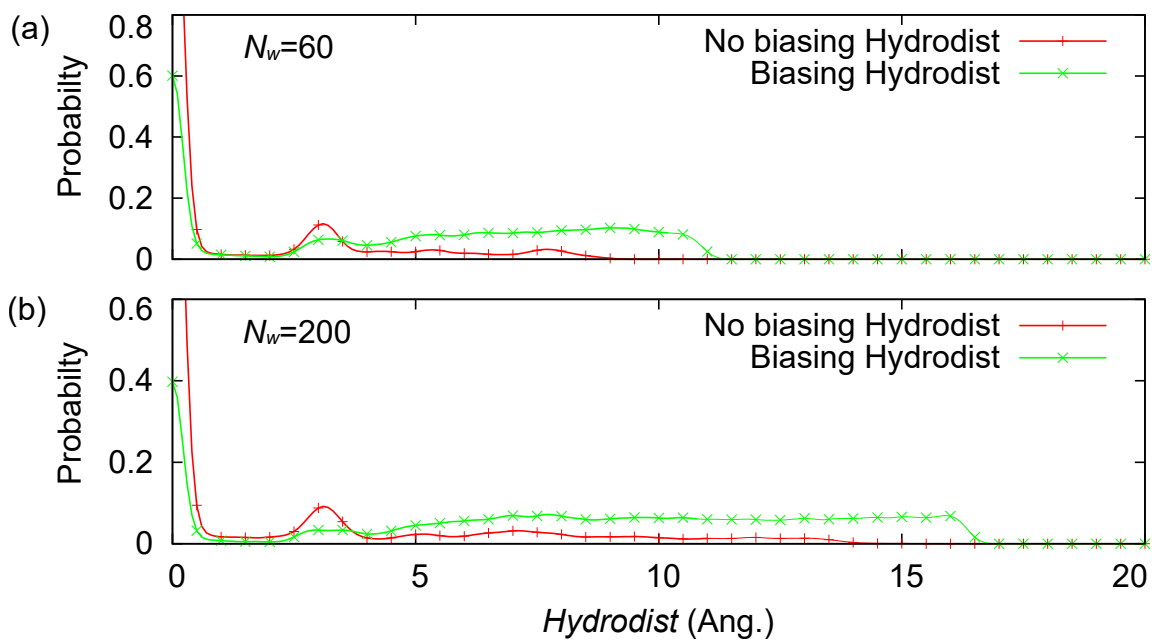

**Figure A3. Histograms of the *Hydrodist* CV. (a)** When *Hydrodist* is biased in the metadynamics simulation. **(b)** when it is not. Histograms were calculated from metadynamics simulations started after 425 ps equilibration. Biasing *Hydrodist* allows a more complete and smoother exploration of the entire simulation volume (see **Fig. A2**).

Clearly, by biasing the *Hydrodist* CV, a more substantial space in both simulation boxes can be explored, and explored more efficiently. This results in better sampling of large separations for the smaller simulation boxes that is needed for accurate $\Delta G$ calculations, and also for a more dynamic and predictable protonation-deportonation cycles for the larger cell. Although 2D metadynamics biasing has added computational costs, it does improve the calculations of $\Delta G$ substantially.